\documentclass[10pt,conference]{IEEEtran}
\IEEEoverridecommandlockouts

\usepackage{cite}
\usepackage{amsmath,amssymb,amsfonts}
\usepackage{algorithmic}
\usepackage{graphicx}
\usepackage{textcomp}
\usepackage{booktabs}
\usepackage{xcolor}
\usepackage{url}
\usepackage{subfig}

\def\BibTeX{{\rm B\kern-.05em{\sc i\kern-.025em b}\kern-.08em
    T\kern-.1667em\lower.7ex\hbox{E}\kern-.125emX}}
\begin{document}

\title{Controllable Singing Style Conversion with Boundary-Aware Information Bottleneck}

\author{
\IEEEauthorblockN{
    Zhetao Hu\textsuperscript{1,2}, 
    Yiquan Zhou\textsuperscript{1,2}, 
    Wenyu Wang\textsuperscript{1,2}, 
    Zhiyu Wu\textsuperscript{3}, 
    Xin Gao\textsuperscript{4}, and 
    Jihua Zhu\textsuperscript{1,2,*}
}

\IEEEauthorblockA{\textsuperscript{1}\textit{School of Software Engineering, Xi'an Jiaotong University}, Xi'an, China}
\IEEEauthorblockA{\textsuperscript{2}\textit{SYKI-SPEECH Team}, Xi'an, China}
\IEEEauthorblockA{\textsuperscript{3}\textit{Fudan University}, Shanghai, China}
\IEEEauthorblockA{\textsuperscript{4}\textit{Division of Music and Audio Union Wheatland Culture and Media Ltd.}, China}
\IEEEauthorblockA{Email: \{huzhetao, zhouyiqian, wenyu.wang\}@stu.xjtu.edu.cn, \\ 
wuzy24@m.fudan.edu.cn, sui@musinya.com, zhujh@xjtu.edu.cn\textsuperscript{*}}
\thanks{* Corresponding author.}
}
\maketitle

\begin{abstract}

This paper presents the submission of the S4 team to the Singing Voice Conversion Challenge 2025 (SVCC2025)—a novel singing style conversion system that advances fine-grained style conversion and control within in-domain settings. To address the critical challenges of style leakage, dynamic rendering, and high-fidelity generation with limited data, we introduce three key innovations: a boundary-aware Whisper bottleneck that pools phoneme-span representations to suppress residual source style while preserving linguistic content; an explicit frame-level technique matrix, enhanced by targeted $F_0$ processing during inference, for stable and distinct dynamic style rendering; and a perceptually motivated high-frequency band completion strategy that leverages an auxiliary standard 48~kHz SVC model to augment the high-frequency spectrum, thereby overcoming data scarcity without overfitting. In the official SVCC2025 subjective evaluation, our system achieves the best naturalness performance among all submissions while maintaining competitive results in speaker similarity and technique control, despite using significantly less extra singing data than other top-performing systems. Audio samples are available online.\footnote{\url{https://csscbaib.netlify.app/}} 

\end{abstract}


\section{Introduction}
Unlike conventional Singing Voice Conversion (SVC)~\cite{zhang2022visinger,9688219,violeta2025singing}, which primarily focuses on timbre or identity mapping, Singing Style Conversion (SSC) aims to finely reshape how a song is performed. This involves modifying vocal techniques such as breathiness, register (falsetto/mixed voice), resonance, vibrato, and glissando, while strictly preserving the lyrical content, melodic structure, and the singer's identity characteristics. In essence, SSC is a style transfer task whose central objective is to rewrite singing techniques and their temporal organization, keeping the underlying content and melody unchanged~\cite{wang2026s}.

Early controllable style transfer studies in speech and singing typically relied on relatively coarse style representations to guide the generator, such as global style embeddings, reference encoders, or utterance-level tokens~\cite{zhang2024stylesinger,huang2022singgan,zhang2022visinger}.
These approaches are effective at capturing global affect or quasi-stationary attributes; However, they often reveal two structural limitations when applied to SSC.
First, many singing techniques exhibit strong locality and rapid temporal dynamics (e.g., time-varying vibrato rates, intermittent breathy segments, or continuous pitch drift in glissando), making it difficult for a single global vector to generate stable and controllable fine-grained variations at the appropriate time steps.
Second, style leakage remains prevalent: prosodic or technique-related cues from the source performance tend to persist in the converted output, compromising the purity of the target style.

\begin{figure*}[t]
    \centering
    \includegraphics[width=\textwidth]{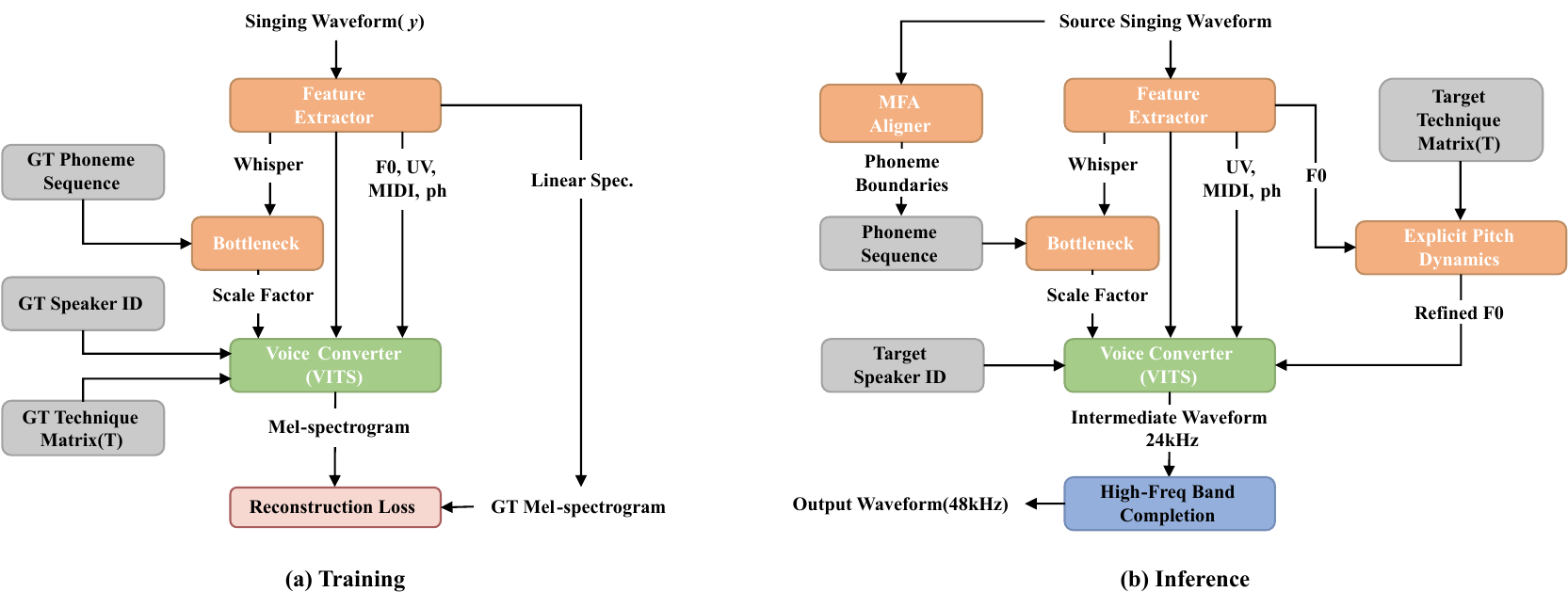}
    \caption{The overall architecture of the proposed System S4. (a) \textbf{Training Stage:} The system employs a boundary-aware semantic bottleneck to suppress style leakage in Whisper features, while the VITS decoder is conditioned on the ground-truth technique matrix $\mathbf{T}$ to learn explicit style rendering. (b) \textbf{Inference Stage:} The pipeline is driven by a target technique matrix, which guides both the explicit pitch dynamics module (for $F_0$ refinement) and the decoder. Finally, a High-Frequency Band Completion module extends the output to 48~kHz fidelity.}
    \label{fig:system_overview}
\end{figure*}

To mitigate these issues, subsequent research has introduced finer-grained supervision and conditioning mechanisms.
A common strategy is to align technique prompts or labels with phoneme- or frame-level time scales, along with richer musical and content conditions (e.g., phonemes, score/MIDI, and $F_0$), to enhance controllability and interpretability~\cite{vyas2023audiobox}.
Recent singing corpora have begun to provide multi-dimensional technique annotations and high-quality score information, enabling explicit technique modeling (e.g., GTSinger~\cite{zhang2024gtsinger}),
and zero-shot singing generation studies further explore hierarchical style control and technique prompting (e.g., TCSinger~\cite{zhang2024tcsinger}).
Nevertheless, style leakage persists.
A key reason is that upstream representations (e.g., self-supervised features~\cite{zhang2026pointcot,zhang2026not,zhang2026cmhanet,zhang2026igasa} or codec tokens) often contain non-content acoustic information from the source signal~\cite{hsu2021hubert,qian2022contentvec,wang2023neural}.
During training and inference, models can exploit such residual cues, causing the target technique rendering to be interfered with or partially overridden by the source style.
Similar prosody leakage has been widely discussed in zero-shot voice conversion and is considered a major factor limiting controllability~\cite{wang2022zero,wang2023lm,zhang2025vevo}.

In recent years, mainstream SSC systems have largely adopted two families of generative backbones.
One family of models employs continuous generative models based on diffusion or flow matching to synthesize mel-spectrograms or waveforms under conditioning~\cite{9688219,ju2024naturalspeech,mehta2024matcha}.
Representative work includes Serenade, which formulates SSC as an audio-infilling problem and applies flow matching, addressing target-style modeling, source-style disentanglement, and melody preservation~\cite{violeta2025serenade}.
The other family employs autoregressive (AR) modeling~\cite{zhang2025ascot,zhang2026chain} over discrete tokens~\cite{zeghidour2021soundstream,wang2023neural} and typically uses a continuous token-to-mel module to recover acoustic details.
This design is common in large-scale speech generation~\cite{anastassiou2024seed,wang2024maskgct,du2024cosyvoice,xie2025towards,touvron2023llama}; for example, VEVO generates content-style token sequences with an AR model and then refines them using a flow-matching acoustic model~\cite{zhang2025vevo,zhang2025vevo2}.
In practice, these two paradigms often appear in a cascaded form (AR for structure and discrete sequences, flow/diffusion for continuous detail), and they therefore share the same key requirement: the conditioning features must strongly disentangle content/melody from style/technique.
If the conditioning features still contain residual source-style information, a powerful generator may amplify this entanglement, resulting in style leakage.
Conversely, if the objective or conditioning constraints are insufficient to capture fast-varying dynamics, the output may exhibit over-smoothing of the technique dynamics.
Recent SVCC2025 evaluations also suggest that, despite substantial progress in naturalness and identity preservation, achieving stable and fine-grained style transfer remains a primary bottleneck~\cite{huang2023singing}.

In this paper, we present System S4, our submission to SVCC 2025 built upon the SYKI-SVC framework~\cite{zhou2025syki}. We address the aforementioned challenges through three targeted designs. First, to mitigate the leakage observed in baselines, we introduce a boundary-aware Whisper bottleneck that pools representations within phoneme spans to wash out residual source style. Second, we employ an explicit frame-level technique matrix enhanced by inference-time $F_0$ processing to ensure the distinct rendering of dynamic styles. Finally, to achieve 48~kHz fidelity without overfitting the limited style data, we propose a high-frequency band completion strategy. Instead of forcing the SSC model to generate the full spectrum, we train a separate, standard 48~kHz SVC model (which is easier to converge) and utilize its output to supplement the high-frequency band ($>10$~kHz) of our style-converted audio.

Subjective evaluations in the SVCC2025 challenge demonstrate that our system achieved the best naturalness performance among all submissions, while maintaining competitive results in both speaker similarity and technique control. Our main contributions are summarized as follows:
\begin{itemize}
    \setlength\itemsep{0em}
    \item \textbf{Semantic bottleneck for disentanglement:} A boundary-aware pooling strategy to suppress style leakage in content representations.
    \item \textbf{Explicit technique control:} A decoupled frame-level conditioning mechanism with targeted $F_0$ processing for dynamic stability.
    \item \textbf{Perceptually motivated high-frequency completion:} A strategy that leverages an auxiliary standard 48~kHz SVC model to complete the high-frequency spectrum, addressing data scarcity challenges in high-fidelity SSC training.
\end{itemize}
The source code is publicly available at\footnote{The implementation is available at \url{https://github.com/HuZhetao/cssc}.}.

\section{Related Work}

Serenade~\cite{violeta2025serenade} is a representative baseline for SVCC2025 and formulates singing style conversion as an audio infilling task: a flow-matching model predicts masked segments of a target mel-spectrogram given the unmasked complement and disentangled features. It also employs cyclic training to disentangle the source style and utilizes a source-filter vocoder with $F_0$ resynthesis to better preserve the melody. This family benefits from expressive modeling and flexible conditioning. However, its inference pipeline is typically heavier than VITS-family systems, and the strict melody constraints may conflict with the desired style-specific pitch dynamics (e.g., vibrato depth or glissando transitions), which can reduce the perceptual salience of dynamic techniques in practice.

Another influential direction uses discrete tokens and a two-stage generation pipeline. Vevo~\cite{zhang2025vevo} proposes the generation of content/style tokens with an autoregressive transformer prompted by a style reference, followed by acoustic generation with a flow-matching transformer prompted by a timbre reference, supported by self-supervised disentanglement using a VQ-VAE bottleneck. The official Vevo1.5 baseline in SVCC2025 further demonstrates the competitiveness of AR+flow pipelines for controllable singing generation. These methods often benefit from large-scale pretraining and carefully designed tokenizers~\cite{anastassiou2024seed,du2024cosyvoice} but they typically
 require greater computational resources and engineering complexity, which may be less favorable under data-limited or latency-constrained settings.

SYKI-SVC builds a high-fidelity SVC system on the SVCC2023 T02 framework, utilizing a VITS-family converter~\cite{zhou2023vits} and a post-processing module~\cite{song2023dspgan} to improve perceptual quality, demonstrating strong naturalness in challenge settings. Nevertheless, in SSC scenarios, commonly used SSL/ASR representations can still retain residual timbre/style information, causing leakage and weakening fine-grained controllability; meanwhile, utterance-level speaker embeddings provide only coarse control and are insufficient for temporally localized technique manipulation. Motivated by these limitations, our submission inherits the efficient, high-naturalness SYKI-SVC backbone and focuses on (i) suppressing content-feature leakage via a boundary-aware Whisper bottleneck and (ii) strengthening the perceptibility of dynamic techniques through explicit frame-level technique conditioning and targeted inference-time pitch dynamics enhancement.

\section{Proposed Method}

We propose a singing style conversion system for SVCC2025, which follows a recognition–synthesis paradigm and consists of three main modules: a feature extraction module, a voice conversion module and a high-frequency band completion module. Given an input singing waveform, the feature extractor derives linguistic and musical conditions, including phoneme sequences, pooled Whisper representations, and pitch-related features such as $F_0$, UV, and MIDI. Based on frame-level technique annotations, a technique matrix is constructed to provide explicit style control. The voice conversion network then generates a 24-kHz singing waveform conditioned on the extracted features, the technique matrix, and the embedding of the target singer. Finally, a super-resolution network supplements high-frequency components above 10 kHz from an auxiliary 48-kHz model, producing the final converted singing voice. 

\begin{figure*}[t]
    \centering
    \includegraphics[width=\textwidth]{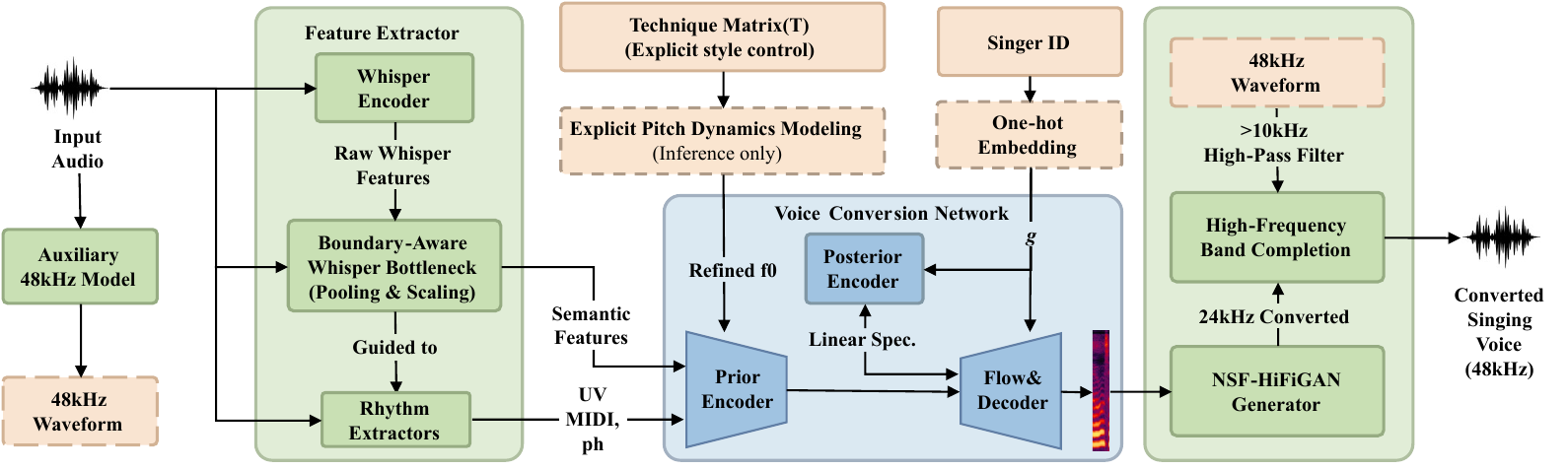}
    \caption{Illustration of the core disentanglement and control mechanisms. The Boundary-Aware Semantic Bottleneck pools frame-level Whisper features within phoneme boundaries to wash out residual source style, while the technique matrix $\mathbf{T}$ explicitly drives style generation.}
    \label{fig:proposed_modules}
\end{figure*}

\subsection{Feature Extraction Module}
Following the SYKI-SVC framework, our system performs singing style conversion via song reconstruction~\cite{zhou2025syki}.
The SVCC2025 dataset provides rich symbolic and acoustic annotations, including phoneme sequences, MIDI notes, $F_0$, and UV flags~\cite{wei2023rmvpe}, which are sufficient to describe lyrics and melody for basic synthesis~\cite{song2022singing}.
To further improve content clarity, we additionally use semantic features from a pretrained Whisper encoder~\cite{radford2023robust} as an auxiliary condition (we also experimented with ContentVec~\cite{qian2022contentvec} and WeNet ASR features~\cite{zhang2022wenet}, but found Whisper yields the best controllability--clarity trade-off under SSC).
Singer identity is conditioned by a one-hot embedding due to the limited number of singers in the challenge set.

\textbf{Phoneme-level Temporal Pooling}
However, we observe that frame-level semantic representations inevitably contain non-semantic information correlated with expressive dynamics, which can introduce technique leakage if directly injected. To mitigate this, we impose a lightweight semantic bottleneck based on phoneme boundaries.
Let $\mathbf{w}_i \in \mathbb{R}^{d}$ denote the raw semantic feature at frame $i$ (e.g., Whisper encoder output).
We use phoneme boundaries to define a phoneme-level neighborhood $\mathcal{N}(t)$ (the set of frames belonging to the same phoneme segment as frame $t$), and average within $\mathcal{N}(t)$:
\begin{equation}
\tilde{\mathbf{w}}_{t}=\frac{1}{|\mathcal{N}(t)|}\sum_{i\in\mathcal{N}(t)}\mathbf{w}_{i}.
\end{equation}
During training, phoneme boundaries are obtained from the provided alignments in the official dataset.
During inference, we use the Montreal Forced Aligner(MFA) ~\cite{mcauliffe2017montreal} trained on the challenge data to obtain phoneme boundaries and construct $\mathcal{N}(t)$.
This pooling suppresses rapid intra-phoneme fluctuations that tend to correlate with expressive techniques, while preserving stable linguistic cues, thereby reducing technique leakage from the semantic branch.
Finally, we apply a global scaling factor $\lambda$ to the pooled features, $\hat{\mathbf{w}}_{t}=\lambda \cdot \tilde{\mathbf{w}}_{t}$, and set $\lambda=0.1$ in all experiments to further limit the dominance of the semantic condition.
\subsection{Voice Converter}

The voice converter is the acoustic synthesis module that recomposes the extracted conditions into a mel-spectrogram while preserving lyrics and melody and enabling explicit technique control. It takes as input the phoneme condition $\text{ph}(t)$, the bottlenecked semantic features $\hat{\mathbf{w}}_t$, and pitch-related features $(F_0(t), \text{UV}(t), \text{MIDI}(t))$, and predicts the target mel-spectrogram $\mathbf{M}$ under the guidance of a technique control signal and a singer identity embedding.

\paragraph{Technique matrix }
To enable time-localized and compositional control over singing techniques, we represent technique annotations as a frame-level binary matrix $\mathbf{T}\in\{0,1\}^{K\times N}$, where $K$ denotes the number of technique labels and $N$ denotes the number of acoustic frames aligned with the mel hop size. Each column $\mathbf{T}_{:,n}$ is a multi-hot vector, allowing multiple techniques to co-occur within the same frame (e.g., breathy+vibrato). During training, $\mathbf{T}$ is constructed by aligning the provided technique segments to the acoustic frame grid. During inference, $\mathbf{T}$ is specified to describe the desired target technique pattern and thus serves as an explicit control interface.

For conditioning, $\mathbf{T}$ is projected by a lightweight embedding layer and concatenated to the frame-level conditioning stream before the prediction of mel. Together with the semantic bottleneck, this explicit injection encourages the converter to attribute the rendering technique to $\mathbf{T}$ rather than implicitly inferring the style from the semantic features, thus improving controllability.

\paragraph{Explicit pitch-dynamics control}
Dynamic techniques such as vibrato and glissando are manifested primarily as fine-grained pitch modulations~\cite{liu2021vibrato}. We find that letting the acoustic model learn these micro-patterns implicitly often leads to temporal over-smoothing. To make pitch-driven techniques more salient and controllable, we refine the contour extracted $F_0(t)$ in the conditioning stage using rule-based transformations derived from $\mathbf{T}$, without introducing an additional predictor $F_0$.
On vibrato-active frames, we inject a sinusoidal modulation in the log-frequency domain:
\begin{equation}
F_0^{\text{vib}}(t)=F_0(t)\cdot 2^{\frac{a \,\sin(2\pi f_v t+\phi)}{1200}},
\label{eq:vib}
\end{equation}
where $a$ is the depth (cents), $f_v$ is the rate (Hz), and $\phi$ is the phase. The modulation is applied only when $m_{\text{vib}}(t)=1$ (from $\mathbf{T}$), otherwise the original contour is kept:
\begin{equation}
F_0^{*}(t)= m_{\text{vib}}(t)\,F_0^{\text{vib}}(t) + \big(1-m_{\text{vib}}(t)\big)\,F_0(t).
\end{equation}

To capture glissando, we replace the discrete pitch shifts at note boundaries with continuous sigmoid transitions. At a boundary time $t_0$ between consecutive notes with target frequencies $F_{prev}$ and $F_{curr}$:
\begin{equation}
F_0^{\text{gliss}}(t)=F_{prev} + (F_{curr}-F_{prev})\cdot \frac{1}{1+e^{-\tau (t-t_0)}},
\label{eq:gliss}
\end{equation}
where $\tau$ controls the transition sharpness. This operation is similarly gated by $m_{\text{gliss}}(t)$ from $\mathbf{T}$ so that transitions are injected only when glissando is desired.

The refined contour $F_0^{*}(t)$ (after applying the technique-gated vibrato/glissando) is then used as pitch conditioning together with $\text{UV}(t)$ and $\text{MIDI}(t)$ for mel generation, improving the salience and temporal consistency of pitch-driven techniques without affecting content or singer identity control.

\subsection{Waveform Generation and High-Frequency Band Completion}
Given the predicted mel-spectrogram $\mathbf{M}$ from the voice converter, we synthesize a 24-kHz waveform using an NSF-HiFiGAN vocoder~\cite{kong2020hifi}. Although 24~kHz is sufficient to model the main-band timbre and technique dynamics, the limited bandwidth often reduces perceived clarity and airiness. A straightforward solution is to directly generate 48-kHz audio; however, the official SVCC dataset is insufficient to train a high-quality 48-kHz vocoder or end-to-end generator. 

To address this, we introduce a lightweight high-frequency band completion strategy using an auxiliary 48-kHz singing conversion model. Empirically, we find that spectral components above 10~kHz contribute mainly to brightness rather than singer identity or technique-related cues. Therefore, during inference we first apply the auxiliary 48-kHz model to the source audio and extract its high-frequency band ($>10$~kHz). We then supplement this band to the converted 24-kHz waveform via smooth frequency-domain mixing, without modifying the low- and mid-frequency content produced by the main system. We denote $\mathcal{H}(\cdot)$ as a high-pass operator extracting $>10$~kHz components, and construct the final waveform as
\begin{equation}
x^{48} = x^{24} + \mathcal{H}\!\left(x^{48}_{src}\right),
\end{equation}
followed by a short cross-fade in the frequency domain to avoid boundary artifacts.This simple band-completion improves perceptual quality while having minimal impact on timbre similarity and technique controllability.

\subsection{Training and Inference Strategy}
This section describes the training and inference rules used to improve disentanglement and technique controllability, without repeating the feature-flow introduced above.

\textbf{Training Strategy.}
We adopt a three-stage protocol to progressively establish audio quality, enforce technique disentanglement, and adapt to singers under the SVCC2025 setting.

\textit{Stage I: Reconstruction Training.}
We first train the main pipeline with standard reconstruction objectives using the official dataset. In this stage, all provided conditions (phoneme/MIDI/$F_0$/UV and semantic features) are enabled to stabilize alignment and establish a strong acoustic foundation.

\textit{Stage II: Technique Disentanglement.}
To reduce technique leakage from the semantic branch, we activate the semantic bottleneck: semantic features are pooled using phoneme boundaries and then down-scaled by a fixed factor (e.g., $\lambda=0.1$). This explicit capacity restriction forces the model to rely on the technique control signal $\mathbf{T}$ for style rendering rather than implicitly encoding techniques in semantic features. During this stage, $\mathbf{T}$ is constructed from ground-truth technique annotations and aligned to the acoustic frame grid, providing supervised technique control for every training sample.

\textit{Stage III: Singer Adaptation Fine-tuning.}
We fine-tune the model to better fit singer-dependent timbre characteristics while preserving the disentangled representations learned in Stage II. In practice, we keep the semantic bottleneck and technique conditioning unchanged and only continue end-to-end optimization on the official dataset.

\textbf{Inference Strategy}
During inference, ground-truth alignments and technique labels are unavailable; therefore, we adopt a rule-based pipeline as follows. We first obtain phoneme boundaries using a self-trained MFA on the challenge dataset, which are required for boundary-aware semantic pooling and feature construction. Next, the desired technique patterns are specified by the target control signal and encoded as a frame-level binary technique matrix $\mathbf{T}$ to condition the converter. For pitch-driven techniques such as vibrato and glissando, we further perform explicit $F_0$ refinement gated by $\mathbf{T}$ prior to mel-spectrogram prediction, so as to enhance perceptual salience and temporal consistency of these rapid dynamics. Finally, to achieve 48-kHz fidelity without coupling it to the main training procedure, the high-frequency band-completion module is trained separately and applied only as a post-processing step after waveform synthesis, remaining outside the primary three-stage training protocol.

\section{Experiments and Results}
\label{sec:experiments}

\subsection{Data Preparation and Training Setup}

All experiments follow the official SVCC2025 \textbf{Task~1} setting; the main 24~kHz system is trained only on the official dataset (about 68 hours) with provided phoneme/MIDI/pitch-related annotations and technique labels for constructing the binary technique matrix $\mathbf{T}$, while the auxiliary 48~kHz model for high-frequency band completion is trained separately and is not involved in the main pipeline.
Training is conducted on 4$\times$RTX~4090 GPUs using AdamW ($\beta_1=0.8, \beta_2=0.99$) with batch size 24, following a three-stage schedule of 200k/300k/100k steps, where Stage~II enables the semantic bottleneck (phoneme-aware pooling and $\lambda{=}0.1$ scaling) to reduce leakage and strengthen reliance on $\mathbf{T}$.

\subsection{Official Results on SVCC2025 Task 1}
The proposed system is evaluated based on the official subjective results of the SVCC2025 challenge~\cite{violeta2025singing}. 
The evaluation framework employs three key metrics: Naturalness (measured via 5-point MOS), and both Singing Style Similarity and Singer Identity Similarity (assessed on a 4-point scale and reported as binary accuracy).

Our submitted system is denoted as \textbf{S4B} (Ours). To verify the efficacy of our disentanglement strategy, we also submitted an ablation system, \textbf{S4A}, which excludes the boundary-aware Whisper pooling and scaling module.

\begin{figure}[htbp]
    \centering
    \includegraphics[width=0.85\linewidth]{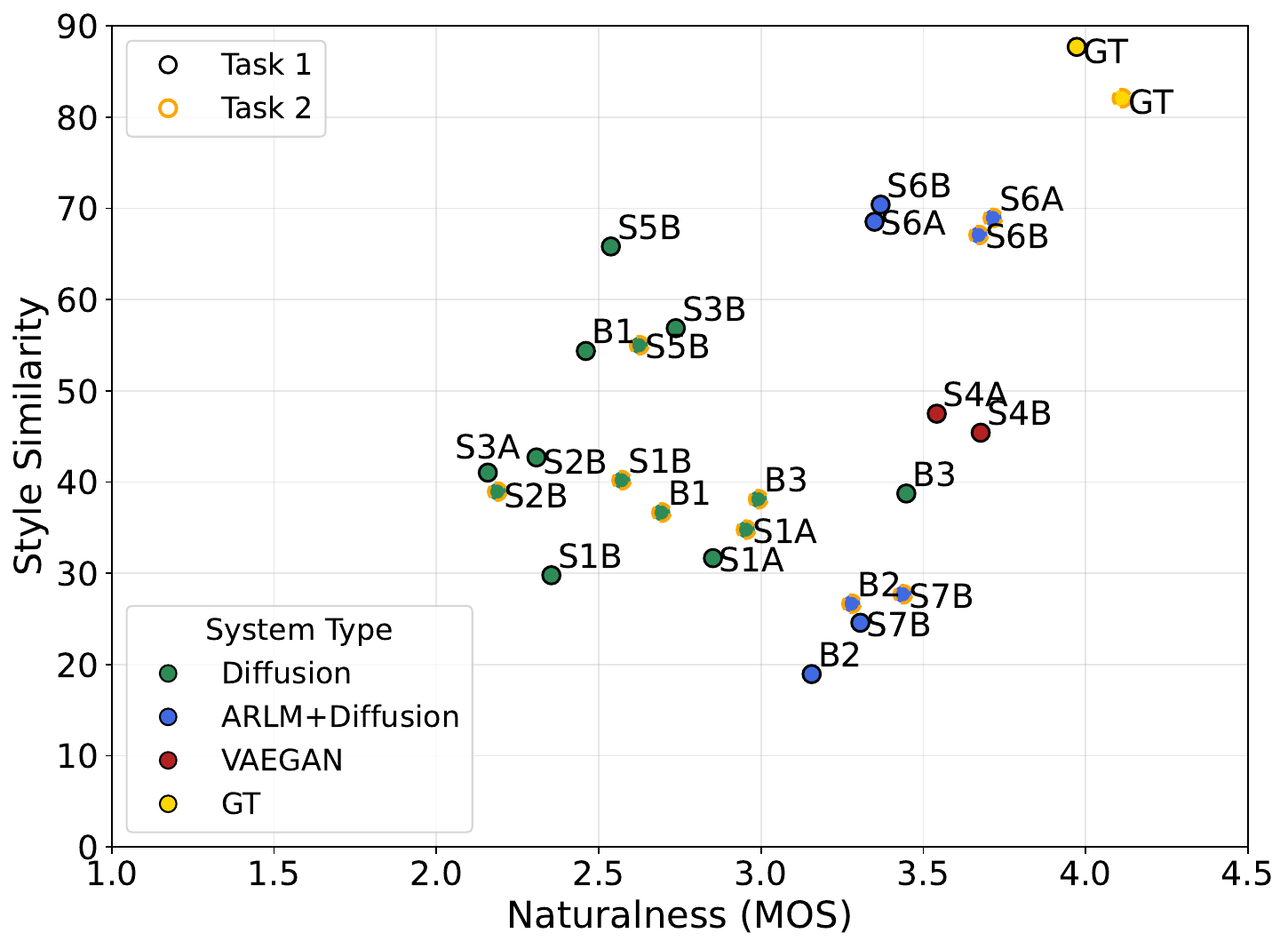} 
    \caption{A scatter plot comparing naturalness and style similarity metrics. Our system S4B is positioned on the far right, indicating superior naturalness performance.}
    \label{fig:scatter_results}
\end{figure}

\begin{figure}[htbp]
    \centering

    \subfloat[Naturalness MOS]{
        \includegraphics[width=0.95\linewidth]{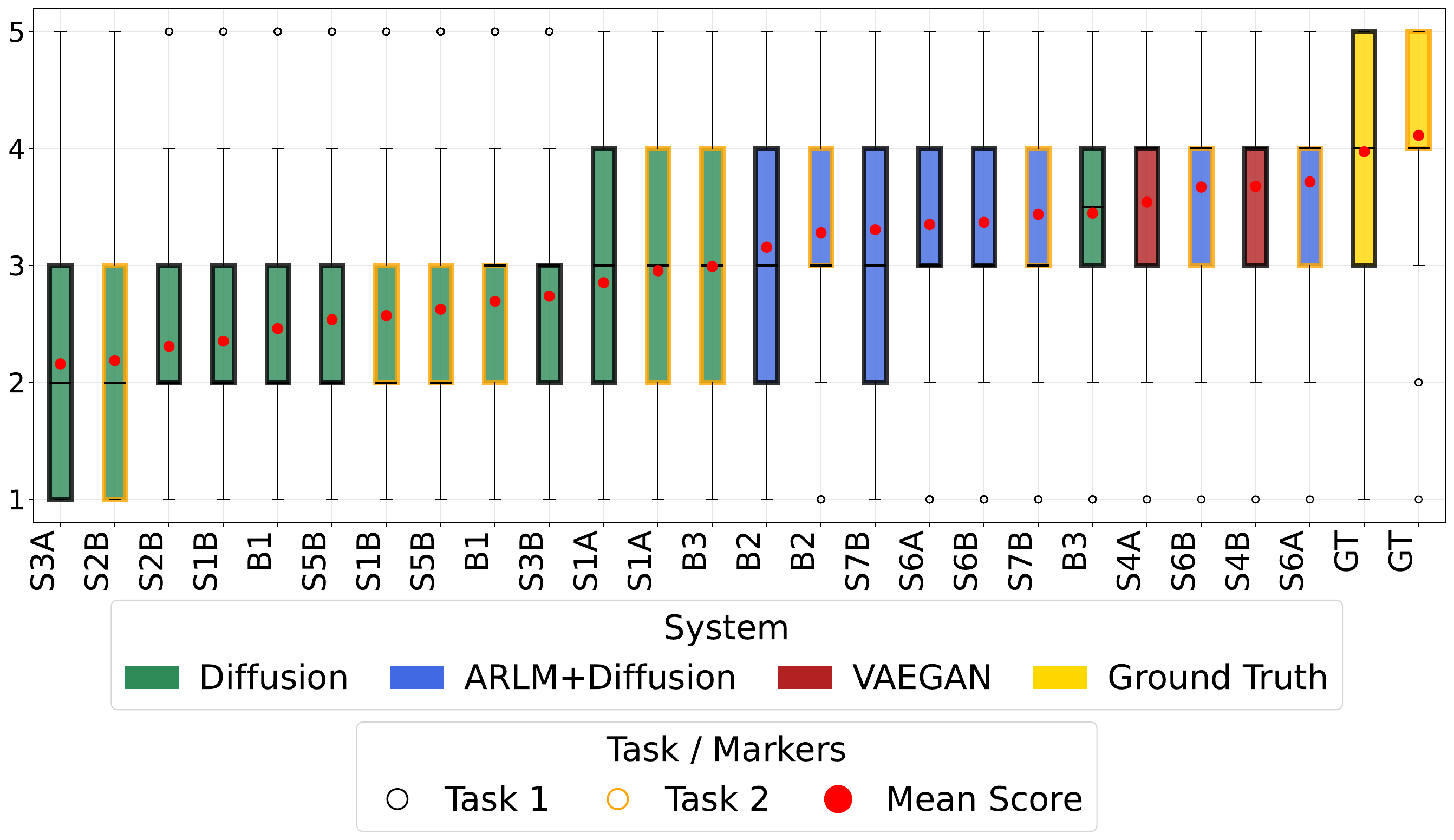}
        \label{fig:mos}
    }

    \vspace{0.5em}

    \subfloat[Singing style similarity]{
        \includegraphics[width=0.95\linewidth]{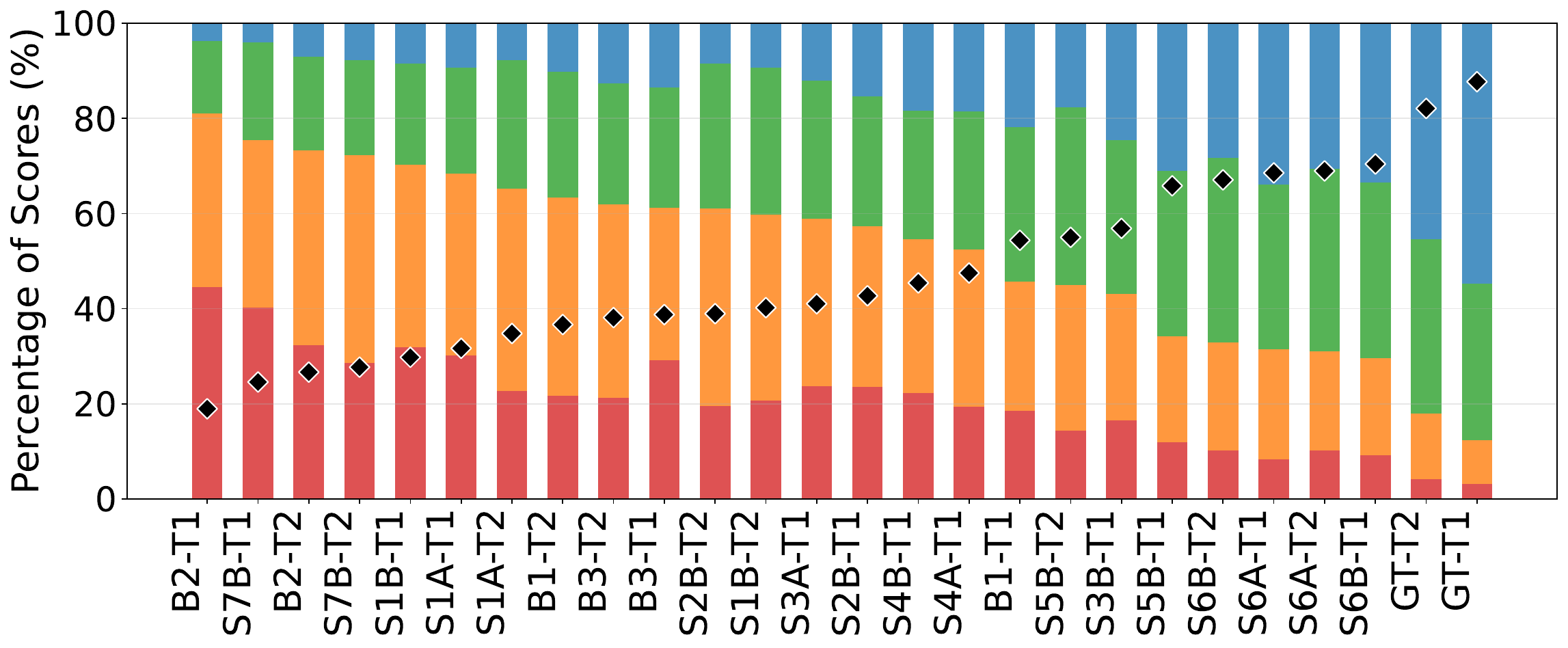}
        \label{fig:style}
    }

    \vspace{0.5em}

    \subfloat[Singer identity similarity]{
        \includegraphics[width=0.95\linewidth]{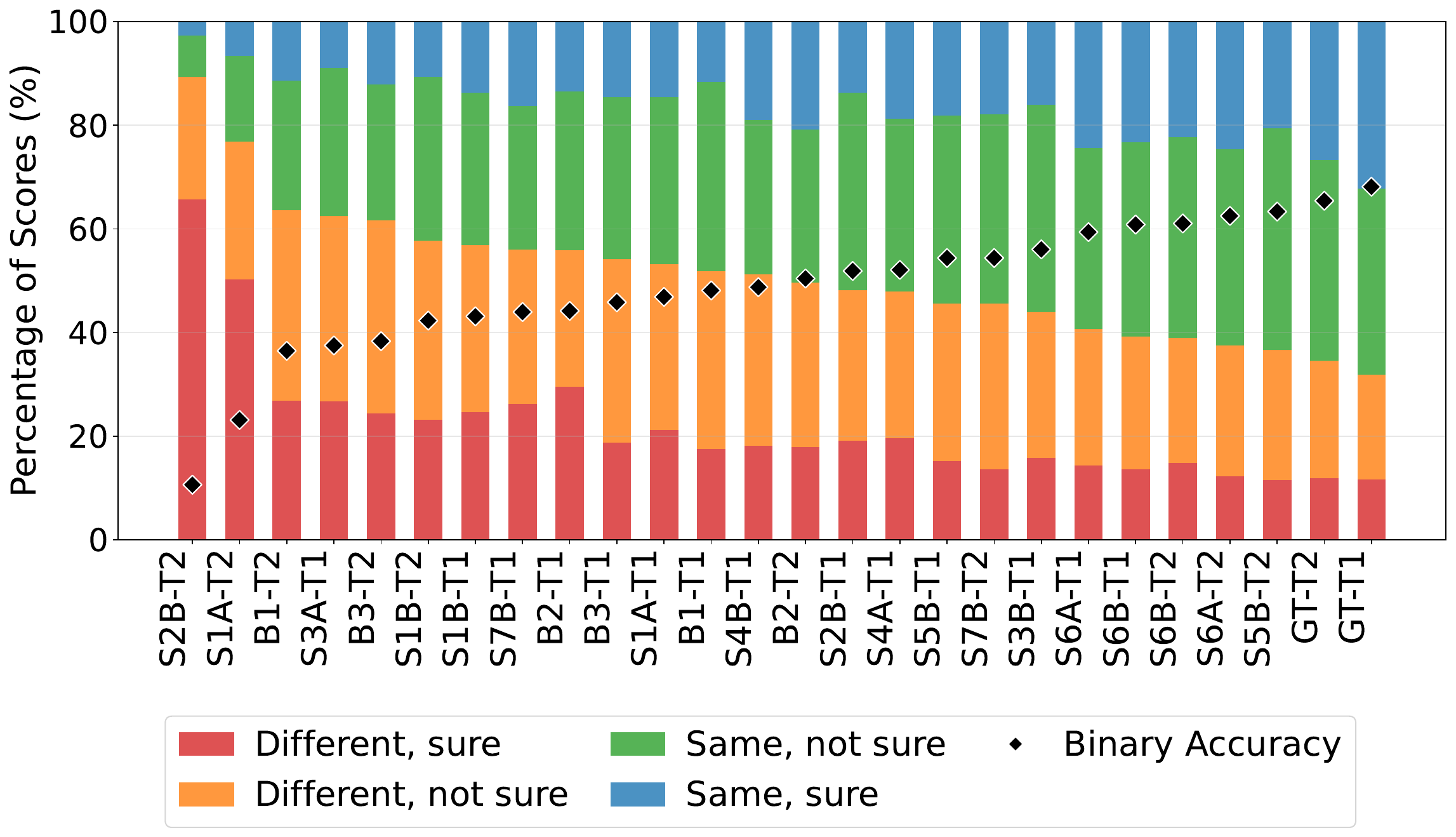}
        \label{fig:identity}
    }

    \caption{Official evaluation results.
    (a) Naturalness MOS boxplots. The red dot indicates the mean score, and the black line indicates the median.
    (b) Singing style similarity.
    (c) Singer identity similarity.
    The black diamonds ($\blacklozenge$) represent the binary accuracy.}
    \label{fig:box_results}
\end{figure}


\subsubsection{Naturalness}
As illustrated in Fig.~\ref{fig:box_results}(a), our system S4B demonstrates exceptional performance in terms of perceptual quality. 
According to the official boxplot results, S4B achieves a median MOS of 4.0, securing the \textbf{1st place in naturalness} among all participating systems (excluding Ground Truth).
The distribution of S4B scores is notably compact and skewed towards the higher end compared to other teams.
This result strongly validates that our explicit technique modeling (via the Technique Matrix) and generative pitch dynamics allow for a highly natural rendition of singing voices, effectively avoiding the robotic artifacts or over-smoothing often observed in conventional SVC systems.

\subsubsection{Similarity and Controllability}
In terms of similarity metrics (Fig.~\ref{fig:box_results}b and c), S4B achieves competitive performance, ranking 4th in style similarity.
While there is a gap in identity similarity compared to the top-ranked systems (e.g., S6, S5), it is crucial to highlight the data efficiency of our approach.
Unlike several top-performing teams that leveraged large-scale external datasets for pre-training to boost timbre disentanglement, our system was trained exclusively on the limited official dataset. 
Given this constraint, our method yields a state-of-the-art naturalness score, proving that the architecture maximizes perceptual quality even under data-limited conditions.

\subsection{Ablation Study}
We evaluate six model variants, each converting the same set of 36 songs. For objective metrics, \textit{Spk Sim} is computed by cosine similarity of CAMPPlus speaker embeddings~\cite{wang2023cam++}, and \textit{MOS} is evaluated by SingMOS~\cite{tang2025singmos}. For human evaluation, listeners rate technique transfer on a 4-level scale (successful / slightly successful / slightly unsuccessful / unsuccessful); we report \textit{Converted} as the proportion of samples rated as successful or slightly successful. The questionnaire also includes a binary option indicating whether the output exhibits \textit{technique leakage}, which is reported as \textit{Leakage}. The six settings cover (i) different semantic features (ContentVec~\cite{qian2022contentvec}, WeNet~\cite{zhang2022wenet}, Whisper~\cite{radford2023robust}), (ii) the effect of scaling ($\lambda$) applied after pooling, and (iii) the effect of removing phoneme-aware pooling.

Table~\ref{tab:ablation} shows that Whisper generally yields better controllability than ContentVec and WeNet. Moreover, the proposed semantic bottleneck (phoneme-aware pooling and $\lambda{=}0.1$ scaling) achieves the best overall trade-off, delivering the highest \textit{Converted} and \textit{Spk Sim} while keeping \textit{Leakage} lowest. In contrast, using an overly strong semantic signal (pooling with $\lambda{=}1.0$) increases leakage, while disabling the semantic signal ($\lambda{=}0$) severely hurts speaker similarity and MOS. Removing pooling also degrades transfer success and increases leakage, indicating that boundary-aware aggregation is important for suppressing style information in the semantic condition.  Finally, regarding the contribution of the high-frequency band completion, we strongly encourage readers to listen to the ablation samples on our demo page.

\begin{table}[htbp]
  \centering
  \caption{Ablation results on semantic features and the proposed semantic bottleneck (phoneme-aware pooling + scaling).}
  \label{tab:ablation}
  \resizebox{\columnwidth}{!}{%
  \begin{tabular}{lcccc}
    \toprule
    \textbf{Setting} & \textbf{Spk Sim} $\uparrow$ & \textbf{MOS} $\uparrow$ & \textbf{Converted} $\uparrow$ & \textbf{Leakage} $\downarrow$ \\
    \midrule
    \textsc{ContentVec} (encoder) & \underline{0.724} & 4.18 & \underline{72.2\%} & 13.9\% \\
    \textsc{WeNet} (encoder)      & 0.694 & 4.04 & 63.9\% & 22.2\% \\
    \midrule
    \textsc{Whisper} + pooling, $\lambda{=}1.0$   & 0.692 & \textbf{4.32} & 66.7\% & 27.8\% \\
    \textsc{Whisper} + pooling, $\lambda{=}0$     & 0.428 & 4.19 & 66.7\% & \underline{11.1\%} \\
    \midrule
    \textsc{Whisper} w/o pooling, $\lambda{=}0.1$ & 0.665 & 4.21 & 58.3\% & 25.0\% \\
    \textsc{Whisper} + pooling, $\lambda{=}0.1$ (ours) & \textbf{0.732} & \underline{4.22} & \textbf{83.3\%} & \textbf{8.3\%} \\
    \bottomrule
  \end{tabular}%
  }
\end{table}

\section{CONCLUSION}
\label{sec:conclusion}
In this paper, we presented System S4 for the SVCC2025 Challenge Task 1.
To achieve controllable singing style conversion under strict data constraints, we proposed a disentanglement strategy that combines a boundary-aware semantic bottleneck with an explicit technique conditioning matrix.
This design effectively suppresses style leakage from content representations, while ensuring precise adherence to target technique controls.
Official evaluation results demonstrate that our system achieves the \textbf{1st place in naturalness} among all participating teams.
The results and ablation studies collectively confirm that our approach offers a data-efficient and high-fidelity solution for singing style conversion.

\bibliographystyle{IEEEtran}
\bibliography{reference}

\end{document}